\documentclass[letterpaper,twocolumn,showpacs,preprintnumbers,nofootinbib,amsmath,amssymb,superscriptaddress]{revtex4}

\newcommand{\ket}[1]{\left\vert#1\right\rangle}
\newcommand{\bra}[1]{\left\langle#1\right\vert}
\newcommand{\ketsub}[2]{\vert#1\rangle_#2}
\newcommand{\brasub}[2]{\!\,_#2\!\langle#1\vert}
\newcommand{\kett}[1]{\vert#1\rangle\!\rangle}
\newcommand{\braa}[1]{\langle\!\langle#1\vert}
\newcommand{\braakett}[2]{\langle\!\langle#1\vert#2\rangle\!\rangle}
\newcommand{\tr}{\textrm{tr}}
\newcommand{\E}{\mathcal{E}}
\newcommand{\R}{\mathcal{R}}
\newcommand{\LL}{\mathcal{L}}
\newcommand{\HH}{\mathcal{H}}
\newcommand{\KK}{\mathcal{K}}
\newcommand{\BB}{\mathcal{B}}

\usepackage{graphicx,amsfonts,epsfig,psfrag}
\begin{document}
\title{Optimum Quantum Error Recovery using Semidefinite Programming}
\author{Andrew S. Fletcher}
\affiliation{Laboratory for Information and Decision Systems, Massachusetts Institute of Technology, 77 Massachusetts Ave. Cambridge, MA 02139}
\affiliation{MIT Lincoln Laboratory, 244 Wood St. Lexington, MA 02420}
\author{Peter W. Shor}
\affiliation{Department of Mathematics, Massachusetts Institute of Technology, 77 Massachusetts Ave. Cambridge, MA 02139}
\author{Moe Z. Win}
\affiliation{Laboratory for Information and Decision Systems, Massachusetts Institute of Technology, 77 Massachusetts Ave. Cambridge, MA 02139}
\date{June 5, 2006}

\begin{abstract}
Quantum error correction (QEC) is an essential element of physical quantum information processing systems.  Most QEC efforts focus on extending classical error correction schemes to the quantum regime.  The input to a noisy system is embedded in a coded subspace, and error recovery is performed via an operation designed to perfectly correct for a set of errors, presumably a large subset of the physical noise process.  In this paper, we examine the choice of recovery operation.  Rather than seeking perfect correction on a subset of errors, we seek a recovery operation to maximize the entanglement fidelity for a given input state and noise model.  In this way, the recovery operation is optimum for the given encoding and noise process.  This optimization is shown to be calculable via a semidefinite program (SDP), a well-established form of convex optimization with efficient algorithms for its solution.    The error recovery operation may also be interpreted as a combining operation following a quantum spreading channel, thus providing a quantum analogy to the classical diversity combining operation.
\end{abstract}

\maketitle
\section{Introduction}

Any implementation of quantum computing or communications requires a strategy for error mitigation.  Indeed, the development of quantum error correction (QEC) schemes was an important early step in moving quantum computing from an interesting theoretical idea to an exciting field with potential for ground-breaking technological implementations\cite{Sho:95}.  The importance of efficient and optimum error mitigation only increases as the field advances.

The earliest efforts  in QEC used encoding techniques modified from classical error correction schemes\cite{Sho:95,Ste:96,CalSho:96,CalRaiShoSlo:97,Got:96}.  Further analysis \cite{KniLaf:97,BenDivSmoWoo:96} laid the foundation for QEC theory, noting that the important metric is how faithful the statistics of the corrected state remain to the ideal behavior.  While that observation suggests quantum error mitigation is thus an optimization problem, most of the subsequent work in the field has appropriately focused on perfect recovery from a set of errors.  This emphasis has allowed many techniques to be borrowed from classical error correction and enabled important feasibility studies in quantum computing.  It is not, however, the only way to consider controlling for quantum errors\cite{LeuNieChuYam:97}.

Recently, some authors \cite{ReiWer:05,YamHarTsu:05} have returned to examining quantum error mitigation as an optimization problem.  The essential properties of a quantum state are the statistics of any observable outcome;  these are completely encapsulated in the density operator $\rho$ of the state.  Noise is introduced by the operation $\mathcal{E}$ which can be thought of as a noisy quantum communications channel.  Thus, the goal of any error correction scheme is to design a recovery operation $\mathcal{R}$  such that the recovered state is as faithful a representation as possible of the input, judged by how well the statistics of observables are preserved.  The optimum recovery minimizes the `distance' between an input density $\rho$ and the output $\R(\E(\rho))$.  This operation may differ from the more traditional QEC recovery operation; such differences illustrate further the contrast between quantum and classical error correction.  To distinguish this approach from QEC, we use the term \emph{quantum error recovery} (QER).  It should be emphasized that the optimum QER recovery operation is dependent on a given input density, encoding operation, and noise model.

The paper is organized as follows.  In Sec. \ref{sec:problem_statement}, we define the parameters for optimum QER.  Section \ref{sec:CPTP} derives a representation of a quantum channel based on a single positive operator.  In section \ref{sec:QEC_SDP}, the optimum recovery operation is cast as a semidefinite program.  Section \ref{sec:div_combining} interprets the recovery operation as an optimum combining problem, and illustrates the computational benefit of such an interpretation.  In Sec. \ref{sec:examples}, QER operations are derived for the amplitude damping channel using codes encoding one qubit into four and five qubits.

\section{Optimum QER}\label{sec:problem_statement}

Most QEC procedures are designed on the principle of `perfect' correction of arbitrary single qubit errors.  Such a design postulates that single qubit errors are the dominant terms in the noise process; thus a scheme that corrects arbitrary single qubit errors and ignores higher order terms will sufficiently mitigate the noise.  Pursuit of this approach has led to important results on the feasibility of quantum error correction, and indeed is a reasonable model for noise processes accurately described by the lower order terms.  However, one may reasonably ask how well this generic approach succeeds in specific cases.

With every quantum code, in the current paradigm, there is an associated recovery operation designed to perfectly correct the dominant errors.  This `traditional' recovery (referred hereafter as the QEC recovery operation) applies a syndrome measurement to determine which error occurred, and a correction operation dependent on the observed syndrome.  For a given code and error process, the QEC recovery operation may not provide the most effective safeguard from error.  Depending upon the form of the error process, an alternate recovery operation may be designed that better preserves the input state, based on some measure of statistical `closeness' between the input density $\rho$ and the output density $\R(\E(\rho))$.  Commonly used metrics\footnote{While not technically a metric, the fidelity is a useful measure of performance for quantum states.} for quantum information arise from the \emph{fidelity}, defined to be
\begin{equation}\label{eq:fidelity}
F(\rho,\sigma)=\tr{\sqrt{\rho^{1/2}\sigma\rho^{1/2}}}
\end{equation}
where $\rho$ and $\sigma$ are density operators.  The fidelity $F$ takes a value between 0 and 1, where 1 indicates that the states are identical.  Using $F_o$ to represent any fidelity-based measure of channel performance, the error recovery optimization problem is to find
\begin{equation}\label{eq:generic_fidelity_max}
\mathcal{R}^\star=\arg\max_\mathcal{R} F_o(\mathcal{R}\circ\mathcal{E}),
\end{equation}
where the maximization is over all valid quantum operations.

We vaguely described the measure of channel performance, $F_o$, as fidelity-based.  In fact, the specific choice of $F_o$ influences the feasibility of the optimization, as well as the interpretation of the result.  There are three main choices of fidelity-based measures of performance, the \emph{minimum fidelity}, the \emph{ensemble average fidelity}, and the \emph{entanglement fidelity}\cite{NieChu:B00}.

The minimum fidelity has the advantage of bounding the performance by seeking the worst case input state.  The metric $F_o$ is the minimization over all inputs of the input-output fidelity.  The optimization becomes two-fold:\footnote{For simplicity of notation, the minimization is shown over all densities $\rho$.  In fact, it is sufficient to minimize over pure state inputs \cite{NieChu:B00}.}
\begin{equation}\label{eq:min_fidelity}
\mathcal{R}^\star=\arg\max_\mathcal{R} \min_\rho F(\rho,\mathcal{R}(\mathcal{E}(\rho))),
\end{equation}
where $F$ is the fidelity given in (\ref{eq:fidelity}).  By virtue of the minimization over $\rho$, one need not assume anything about the input state.  This was the metric of choice in \cite{KniLaf:97} first establishing the theory of QEC.  The disadvantage arises through the complexity of the metric; indeed the optimization problem is now over two sets, and the metric $F_o$ is not linear with respect to $\R$.  Efficient routines that have been developed for solving the optimization problem of (\ref{eq:min_fidelity}) are sub-optimum\cite{YamHarTsu:05}.

We obtain a more tractable optimization problem if we are willing to assume some form for the input distribution $\rho$, particularly if the metric $F_o$ can be reasonably defined as linear in the recovery operation $\R$.  While assuming a form for $\rho$ makes the solution less general, it illuminates important characteristics of quantum error mitigation by enabling construction of an optimum recovery operation.  Given a value of $\rho$, we may use either the ensemble average fidelity or the entanglement fidelity.  

Ensemble average fidelity models the input as being in a state $\rho_i$ with probability $p_i$.  We define an arbitrary quantum channel $\BB$ on the Hilbert space $\HH$ as $\BB:\LL(\HH)\mapsto\LL(\HH)$ where $\LL(\HH)$ indicates the set of bounded linear operators on $\HH$.  The measure by which $\BB$ preserves the input is then the average squared fidelity:
\begin{equation}
\bar{F}(p_i,\rho_i,\BB)=\sum_ip_iF(\rho_i,\BB(\rho_i))^2.
\end{equation}
$\bar{F}$ is linear in $\BB$ when each $\rho_i$ describes a pure state; for linearity we must assume more than just the density of the input.

Entanglement fidelity\cite{Sch:96} arises from the mathematical concept of mixed state purification.  Any mixed quantum state can be represented as a subsystem of a pure state in a larger Hilbert space.  The subsystem is mixed due to the entangled nature of the pure state.  Consider a mixed state $\rho\in\LL(\HH)$.  By defining a reference space $\mathcal{A}$, we may denote $\rho$ as a subsystem of a pure state:
\begin{equation}
\rho=\tr_\mathcal{A}\ket{AH}\bra{AH}
\end{equation}
 where $\ket{AH}$ is a pure state in the space $\mathcal{A}\otimes\HH$.  The entanglement fidelity then measures how faithfully $\BB$ maintains the purification (or equivalently, how well it preserves the entanglement). It is given by\footnote{We use $F$ to signify both the fidelity and the entanglement fidelity.  The distinction should be obvious from context, as the arguments for the fidelity are two density operators, whereas for the entanglement fidelity, the arguments are a density operator and a channel mapping.}
\begin{equation}\label{eq:ent_fid}
F(\rho,\BB)=\bra{AH}\mathcal{I}_\mathcal{A}\otimes\BB(\ket{AH}\bra{AH})\ket{AH},
\end{equation}
the squared fidelity of the input $\ket{AH}$ with the output of the channel $\mathcal{I}_\mathcal{A}\otimes\BB$.

With entanglement fidelity as $F_o$ in  (\ref{eq:generic_fidelity_max}), we define the optimum recovery operation for the error process $\E$ and input distribution $\rho$ as
\begin{equation}\label{eq:ent_fidelity_max}
\mathcal{R}^\star_\rho=\arg\max_\mathcal{R} F(\rho,\mathcal{R}\circ\mathcal{E}).
\end{equation}

\section{CPTP Maps and Positive Operators}\label{sec:CPTP}

A valid quantum operation must be completely positive (CP) and trace-preserving (TP)\cite{Kra:B83}.  This requirement follows directly from the postulates of quantum mechanics wherein the evolution of a closed quantum system is unitary.  
Let $\E$ be a CPTP map from $\LL(\HH)\mapsto\LL(\KK)$.  The most familiar representation of the CPTP map is the \emph{Kraus} (or operator sum) form, where the mapping is specified by a set of operators $\{E_k\}$ known as the operator elements\cite{Kra:B83}.  The channel output is given by
\begin{equation}\label{eq:Kraus}
\E(\rho)=\sum_k E_k \rho E_k^\dagger,
\end{equation}
and the CPTP constraint is met when
\begin{equation}
\sum_k E_k^\dagger E_k = I,
\end{equation}
where $I$ is the identity operation on $\LL(\HH)$.  

While a properly constrained set of operators fully specify a quantum channel, the Kraus form is an inconvenient one for optimization.  The most obvious inconvenience is the many-to-one correspondence between sets of operator elements and channel mappings.  For this reason, we will utilize the representation of a channel mapping by a positive semidefinite operator on $\LL(\KK\otimes\HH)$, which correspondence is one-to-one \cite{DarLop:01}.  We will refer to the operator description of CPTP maps as the \emph{superoperator} form\cite{Cav:99,Dep:67}.

The superoperator form may be derived by recognizing that the space of bounded linear operators forms a Hilbert space.  It is convenient to have a general method to convert the operator to a ket notation.  Let $C$ be a bounded linear operator from $\HH_2$ to $\HH_1$: $C\in\LL(\HH_2,\HH_1)$.  We define a ket in the Hilbert space $\HH_1\otimes\HH_2$ associated with $C$ as\footnote{The notation $\kett{\cdot}$ is used to emphasize that these kets represents operators, not quantum states.}
\begin{equation}
\kett{C}=\sum_{ij} c_{ij}\ketsub{i}{1}\ketsub{j}{2}.
\end{equation}
where $\{\ketsub{i}{1}\}$ and $\{\ketsub{j}{2}\}$ are orthonormal bases for $\HH_1$ and $\HH_2$, respectively, and $c_{ij}=\brasub{i}{1}C\ketsub{j}{2}$ is the matrix element of $C$ on these bases.

Two useful relations follow directly from this definition.  The first one,
\begin{equation}
A\otimes B\kett{C} = \kett{ACB^T}\label{eq:kett_triple_product},
\end{equation}
applies whenever the dimensions of $A$, $B$, and $C$ indicate that $ACB^T$ is a valid operator.\footnote{Note the symbol $^T$ indicates the transpose with respect to the specified basis \emph{without conjugation}.}  The second relation applies for $C_1, C_2\in\LL(\HH_2,\HH_1)$:
\begin{equation}
\tr_{\HH_2}[\kett{C_1}\braa{C_2}]=C_1C_2^\dagger \in \LL(\HH_1).\label{eq:kett_trace}
\end{equation}

From these relations, we see that the channel mapping $\E:\LL(\HH)\mapsto\LL(\KK)$ can be given by
\begin{eqnarray}
\E(\rho)&=&\sum_k E_k \rho E_k^\dagger\nonumber\\
\nonumber&=& \sum_k \tr_\HH \left [\kett{E_k\rho}\braa{E_k}\right ]\\
\nonumber&=& \sum_k \tr_\HH \left [I\otimes\rho^T\kett{E_k}\braa{E_k}\right ]\\
&=& \tr_\HH \left [I\otimes\rho^T X_\E\right ],\label{eq:Pos_Op_channel}
\end{eqnarray}
where $X_\E\equiv\sum_k\kett{E_k}\braa{E_k}$.
The trace-preserving property that
\begin{equation}
\tr_\KK\E(\rho)=1=\tr_\HH[\rho^T\tr_\KK[X_\E]]
\end{equation}
for all density operators $\rho\in\LL(\HH)$ can be stated as
\begin{equation}
\tr_\KK X_\E=I\in\LL(\HH).
\end{equation}
In the superoperator form, the entire mapping $\E$ is specified by the positive operator $X_\E$.

\section{Optimum Recovery via Semidefinite Programming}\label{sec:QEC_SDP}

The problem given by  (\ref{eq:ent_fidelity_max}) is a convex optimization problem and we may approach it with sophisticated tools.  Particularly powerful is the semidefinite program (SDP) \cite{VanBoy:96}, where the objective function is linear in an input constrained to a semidefinite cone.  Indeed, the power of the SDP was a primary motivation in choosing to maximize the entanglement fidelity, which is linear in the quantum operation $\R$.

The definition of entanglement fidelity given in  (\ref{eq:ent_fid}) is intuitively useful, but awkward for calculations.  An easier form arises when operator elements $\{B_i\}$ for $\BB$ are given.  The entanglement fidelity is then 
\begin{equation}\label{eq:ent_fid_kraus}
F(\rho,\BB)=\sum_i|\tr(\rho B_i)|^2.
\end{equation}
From  (\ref{eq:ent_fid_kraus}), we may derive a calculation rule for the entanglement fidelity when the channel $\BB$ is expressed as in the superoperator form.  Using  (\ref{eq:kett_trace}), we see that $\tr{B_i \rho=\tr{\kett{B_i}\braa{\rho}}=\braakett{\rho}{B_i}}$.  Inserting this into  (\ref{eq:ent_fid_kraus}), we obtain the entanglement fidelity in terms of $X_\BB$:
\begin{eqnarray}
\nonumber F(\rho,\BB)&=&\sum_i\braakett{\rho}{B_i}\braakett{B_i}{\rho}\\
&=&\braa{\rho}X_\BB\kett{\rho}.
\end{eqnarray}

Armed with this expression for the entanglement fidelity, we may now express  (\ref{eq:ent_fidelity_max}) in a form readily seen to be a semidefinite program.  To do this, we must consider the form of the composite channel $\R\circ\E:\LL(\HH)\mapsto\LL(\HH)$ expressed as a positive operator on $\HH\otimes\HH$.  If the operator elements for each channel are $\{R_i\}$ and $\{E_j\}$, then the operator $X_{\R\E}$ is given by
\begin{equation}
X_{\R\E}=\sum_{ij}\kett{R_iE_j}\braa{R_iE_j}.
\end{equation}
Applying  (\ref{eq:kett_triple_product}), this becomes
\begin{eqnarray}
\nonumber X_{\R\E}&=&\sum_{ij}I\otimes E_j^T\kett{R_i}\braa{R_i}I\otimes E_j^*\\
&=& \sum_j (I\otimes E_j^T)X_\R (I\otimes E_j^*),
\end{eqnarray}
where the $^*$ represents complex conjugation, without transposition.  The entanglement fidelity is then
\begin{eqnarray}
\nonumber
F(\rho,\R\circ\E)&=&\sum_j \braa{\rho}(I\otimes E_j^T)X_\R (I\otimes E_j^*)\kett{\rho}\\
&=& \tr{X_\R C_{\rho,\E}}
\end{eqnarray}
where
\begin{eqnarray}\nonumber
C_{\rho,\E}&=&\sum_j I\otimes E_j^*\kett{\rho}\braa{\rho}I\otimes E_j^T\\
&=& \sum_j \kett{\rho E_j^\dagger}\braa{\rho E_j^\dagger}.
\end{eqnarray}

We may now express the optimization problem  (\ref{eq:ent_fidelity_max}) in the simple form
\begin{eqnarray}\nonumber
X_{\R_{\rho}}^\star=\arg\max_X \tr ({X C_{\rho,\E}})\\ \textrm{such that }
X \geq 0, \hspace{10 pt} \tr_\HH{X}=I.\label{eq:ent_fid_max}
\end{eqnarray}
This form illustrates plainly the linearity of the objective function and the semidefinite and equality structure of the constraints.  Indeed, this is the exact form of the optimization problem in \cite{AudDem:02} which first pointed out the value of semidefinite programs (SDP) for optimizing quantum channels.

The value of an SDP for optimization is two-fold.  First, an SDP is a sub-class of convex optimization, and thus a local optimum is guaranteed to be a global optimum.  Second, there are efficient and well-understood algorithms for computing the optimum of a semidefinite program.  These algorithms are sufficiently mature to be widely available.  Thus, by expressing the optimum recovery channel as an SDP, we have explicit means to compute the solution numerically.


\section{Quantum Diversity Combining}\label{sec:div_combining}

In the preceding analysis, the method of encoding has been implied by the choice of $\rho$.  Indeed, in most treatments of QEC the input density is restricted to a subspace called the quantum error correcting code (QECC).  If $P_C$ is a projection operator onto the code subspace, then $\rho=P_C\rho P_C$ implies that the state $\rho$ is within the code subspace.  The channel is typically defined such that the input and output spaces are identical (\emph{i.e.} $\HH=\KK$) and the noise process generally perturbs the system from the code subspace.  While this representation is a perfectly legitimate model for the error process, and convenient when viewing QEC in a mode comparable to classical error correction, the dimensionality is unnecessarily high for the optimization routine.

Recall that $X_\R$ is a Hermitian operator on the space $\KK\otimes\HH$.  The optimization thus has $d_\HH^2 d_\KK^2$ real parameters.  Even for a $[[5,1,3]]$\footnote{The notation $[[n,k,d]]$ refers to a quantum code encoding $k$ qubits of information into a $n$ qubit system.  The parameter $d$ refers to the \emph{weight} of the code\cite{NieChu:B00}.} code, the smallest for arbitrary single qubit errors\cite{BenDivSmoWoo:96,LafMiqPazZur:96}, this optimization then has $2^{20}$ optimization variables.  The high dimensionality can be alleviated somewhat by embedding the encoding into the noise process, and redefining $\E$ as a \emph{quantum spreading channel} where $d_\HH<d_\KK$.
\begin{figure}
\begin{center}


\includegraphics[width=\columnwidth]{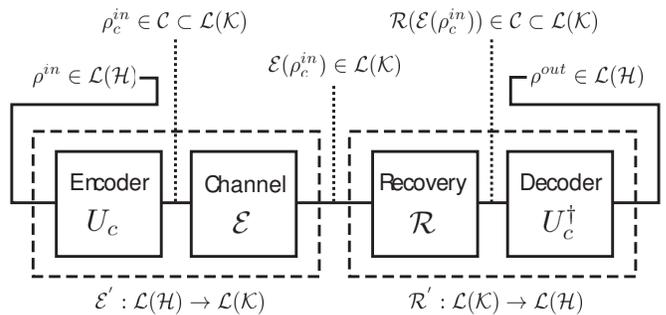}

\end{center}

\caption{Transform from standard quantum error correction to quantum spreading channel.  By considering the error channel $\E$ together with the encoding $U_C$ and the recovery operation $\R$ with the decoder $U_C^\dagger$, the dimension of the SDP may be reduced by a factor of $(\dim{\KK}/\dim{\HH})^2$.  Note that the output of the encoder $U_C$ and the recovery channel $\R$ live on the code subspace $\mathcal{C}$.}\label{fig:spreading_channel}
\end{figure}

The transform to a quantum spreading channel is illustrated in Fig. \ref{fig:spreading_channel}.  Consider the noise process $\E:\KK\mapsto\KK$ with operator elements $\{E_i\}$, input density $\rho\in\LL(\KK)$, and code projector 
\begin{equation}
P_C=\sum_n^{d_\HH}\ketsub{n}{L}\brasub{n}{L}
\end{equation}
where $\ket{n}_L\in\KK$ are the logical states of the code.  Since the input is in the codespace, $\rho$ is preserved by the code projector: $\rho=P_C\rho P_C$.  We can reduce the dimensionality of the optimization by transforming this problem.  Let $\HH$ be a $d_\HH$ dimensional Hilbert space with orthonormal basis $\{\ket{n}_\HH\}$.  $\HH$ can be considered the space of the information source.  The encoding process is accomplished by the operator $U_C=\sum_n^{d_\HH}\ketsub{n}{L}\brasub{n}{\HH}$ that maps the basis states of $\HH$ to the logical states in $\KK$.  $U_C$ is an isometry: $U_C^\dagger U_C=I$.  Note that
\begin{eqnarray}
\nonumber U_CU_C^\dagger&=&\sum_n^{d_\HH}\ketsub{n}{L}\brasub{n}{L}\\
&=&P_C.
\end{eqnarray}
If we redefine $\rho'=U_C^\dagger\rho U_C\in\LL(\HH)$ and the operator elements $E_i'=E_i U_C$, then we see the processes $\E(\rho)$ and $\E'(\rho')$ are identical:
\begin{eqnarray}\nonumber
\E'(\rho')&=&\sum_i E_i U_C U_C^\dagger \rho U_C U_C^\dagger E_i^\dagger\\
\nonumber&=&\sum_i E_i P_C \rho P_C E_i^\dagger\\
&=&\E(\rho).
\end{eqnarray}
By enacting such a transformation, the optimization dimension is reduced from $d_\KK^4$ to $d_\KK^2d_\HH^2$.  For the $[[5,1,3]]$ code, the reduction is from $2^{20}$ to $2^{12}$.

The above transformation illustrates an alternative interpretation of recovering an encoded quantum state  after an error process.  We may instead consider an unencoded state input into a \emph{quantum spreading channel}, a channel in which the output dimension is greater than the input dimension (\emph{i.e.} $\dim{\KK}>\dim{\HH}$).  The recovery operation is an attempt to combine the spread output  back into the input space, presumably with the intent to minimize information loss.  The recovery operation is then the quantum analog to the classical communications concept of diversity combining.

Classical diversity combining describes a broad class of problems in communications and radar systems.  In its most general form, we may consider any class of transmission problems for which the receiver observes multiple transmission channels.  These channels could arise due to multi-path scattering, frequency diversity (high bandwidth transmissions where channel response varies with frequency), spatial diversity (free-space propagation to multiple physically separated antennas),  time diversity, or some combination of the four.  Diversity combining is a catch-all phrase for the process of exploiting the multiple channel outputs to improve the quality of transmission (\emph{i.e.} by reducing error or increasing data throughput).

On its face, an extension of diversity combining to the quantum communications regime is unclear.  Indeed, the simplest way to understand classical diversity - receiving multiple copies of the input, independently affected by the channel - might also lead one to conclude that the concept is not applicable to quantum communication.  After all, the fundamental postulates of quantum mechanics lead to the ``no-cloning'' theorem; multiple, separable copies of the input state are ruled out by theorem.  A more careful consideration of classical diversity combining, however, illuminates the parallels between the quantum and classical viewpoints.

In a general description of classical diversity, the input signal is coupled through the channel to a receiver system of higher dimensionality.  Consider a communication signal with a single input antenna and $N$ receive antennae.  Often, the desired output is a signal of the same dimension as the input, a scalar in this case.  Diversity combining is then the process of extracting the maximum information from the $N$-dimensional received system.  In most communications systems, this combining is done at either the analog (leading to beam-forming or multi-user detection) or digital (making the diversity system a kind of repeater code) levels.  Thus, the natural inclination is to equate diversity combining with either beam-forming or repeater codes.  The most general picture, however, is that of recombining the received signal that was spread by the channel from the input system to a higher dimensional output system.  Thus, it is appropriate to consider a quantum spreading channel to be quantum diversity channel, and the recovery operation is a quantum diversity combiner.

\section{Examples}\label{sec:examples}

To illustrate the potential benefit of optimum QER, we numerically compare the performance of the optimum QER with QEC recovery for two encoding schemes.  First, we examine the $[[5,1,3]]$ code in the presence of the amplitude damping channel.  Second, we compare the 4-bit amplitude damping code from \cite{LeuNieChuYam:97} with the optimum QER.  The latter is a particularly apt choice for comparison, as the code was designed for a specific channel and sought to only approximately satisfy the quantum error correcting conditions.

In both cases, the noise channel $\E_a$ is the amplitude damping channel, which for a single qubit has operator elements
\begin{equation}\label{eq:ampdamp}
E_0=\left [ \begin{array}{ccc} 1 & 0 \\ 0 &\sqrt{1-\gamma} \end{array} \right ]\hspace{.5 cm} \textrm{and} \hspace{.5 cm}
E_1=\left [ \begin{array}{ccc} 0 & \sqrt{\gamma} \\ 0 & 0 \end{array} \right ].
\end{equation}
This channel is a commonly encountered model, where the parameter $\gamma$ indicates the probability of decaying from state $\ket{1}$ to $\ket{0}$ (\emph{i.e.} the probability of losing a photon).  Amplitude damping is a logical choice to illustrate the benefits of optimum QER as the operation is not symmetric with respect to $\ket{0}$ and $\ket{1}$.

\subsection{Five-Qubit Code}

The five-qubit code was independently discovered by \cite{BenDivSmoWoo:96} and \cite{LafMiqPazZur:96}.  We will here follow the treatment in \cite{NieChu:B00}, and specify the code via the generators $\{g_1,g_2,g_3,g_4\}$ and the logical $\bar{Z}$ and $\bar{X}$ operations specified in Table \ref{table:5qubitcode}.  The code subspace $\mathcal{C}$ is the two-dimensional subspace that is the $+1$ eigenspace of the generators $g_i$.  The logical states $\ketsub{0}{L}$ and $\ketsub{1}{L}$ are the $+1$ and $-1$ eigenkets of $\bar{Z}$ on $\mathcal{C}$.  

\begin{table}[tbh]\label{table:5qubitcode}
\begin{tabular}{c|c}
\hline
\hline
Name & Operator\\
\hline
$g_1$ & $X Z Z X I$\\
$g_2$ & $I X Z Z X$\\
$g_3$ & $X I X Z Z$\\
$g_4$ & $Z X I X Z$\\
$\bar{Z}$ & $Z Z Z Z Z$\\
$\bar{X}$ & $X X X X X$\\
\hline
\hline
\end{tabular}
\caption{Generators and logical operations of the five qubit (\emph{i.e.} $[[5,1,3]]$) code.}
\end{table}

\begin{figure}[hbt]
\begin{center}
{\includegraphics[width=\columnwidth]{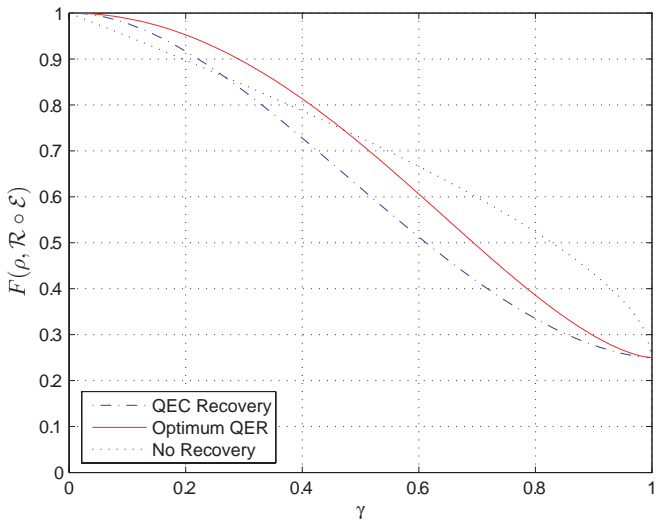}}
\end{center}
\caption{Entanglement fidelity vs.\ $\gamma$ for the 5 qubit code and the amplitude damping channel $\E_a^{\otimes 5}$.  $\gamma$ refers to the damping parameter of the channel.  Entanglement fidelity with no recovery operation is included for comparative purposes.}\label{fig:5qubit_comp}
\end{figure}

To compute the optimum recovery for this code, we assume that the logical states are equally likely, that is $\rho=\frac{1}{2}\ketsub{0}{L}\brasub{0}{L}+\frac{1}{2}\ketsub{1}{L}\brasub{1}{L}$.  (This choice of $\rho$ in fact assumes nothing about the choice of codewords; rather it is the maximum entropy distribution constrained to the code space $\mathcal{C}$.)  The noise channel is $\E_a^{\otimes 5}$, the amplitude damping channel acting independently on each qubit.  For each choice of the parameter $\gamma$, the optimum recovery operation $\R^\star_\rho$ is computing according to (\ref{eq:ent_fidelity_max}).  We compare the entanglement fidelity $F(\rho,\R\circ\E_a^{\otimes 5})$ for both $\R^\star_\rho$ and $\R_{\textrm{QEC}}$ in fig. \ref{fig:5qubit_comp}.  

Figure \ref{fig:5qubit_comp} illustrates clearly the difference between optimum QER and QEC recoveries for large values of $\gamma$.  It is also instructive to compare the techniques for small values of $\gamma$.  We do this numerically by calculating the polynomial expansion of $F(\rho,\R\circ\E)$ as $\gamma$ goes to zero.  The entanglement fidelity for the optimum QER has the form $F(\rho,\R\circ\E)\approx 1 -1.166 \gamma^2+\mathcal{O}(\gamma^3)$.  In contrast, the QEC recovery is $F(\rho,\R\circ\E)\approx 1 -2.5 \gamma^2+\mathcal{O}(\gamma^3)$.

\subsection{Four-Qubit `Approximate' Code}

In \cite{LeuNieChuYam:97}, Leung \emph{et. al.} recognized the advantage that channel-specific error recovery schemes can have over generic QEC routines.  To illustrate the advantage, they designed a code specific to the amplitude damping channel with good performance for small $\gamma$ that only required 4 qubits of overhead, in contrast to the generic five qubit code.  The code only approximately meets the QEC conditions, and as a result, the `corrected' state is somewhat distorted from the input, even when the dominant error term occurs.  In this way, the procedure is based upon principles similar to the optimum QER we have developed.

\begin{figure}[hbt]
\begin{center}
{\includegraphics[width=\columnwidth]{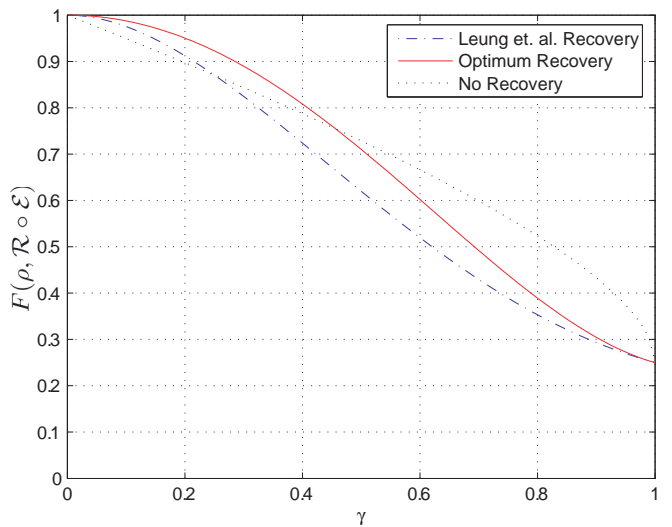}}
\end{center}
\caption{Entanglement fidelity vs.\ $\gamma$ for the 4 qubit amplitude damping code of Leung \emph{et.\! al.\! } \cite{LeuNieChuYam:97}.}\label{fig:Leung_comp}
\end{figure}

The logical states are given by
\begin{eqnarray}
\ketsub{0}{L}= \frac{1}{\sqrt{2}}(\ket{0000}+\ket{1111})\\
\ketsub{1}{L}= \frac{1}{\sqrt{2}}(\ket{0011}+\ket{1100}),
\end{eqnarray}
and the recovery operation is specified by the circuits in Figure 2 of \cite{LeuNieChuYam:97}.  We note that the recovery operation depends explicitly on the parameter $\gamma$.  We compare the recovery of Leung \emph{et. al.} with the optimum QER computed according to  (\ref{eq:ent_fidelity_max}), once again assuming the completely mixed input density $\rho=\frac{1}{2}\ketsub{0}{L}\brasub{0}{L}+\frac{1}{2}\ketsub{1}{L}\brasub{1}{L}$.  The numerical comparison, for various values of $\gamma$ is provided in fig. \ref{fig:Leung_comp}.  As $\gamma$ goes to zero, the entanglement fidelity for the optimum QER is numerically determined to have the form $F(\rho,\R\circ\E)\approx 1 -1.25 \gamma^2+\mathcal{O}(\gamma^3)$.  In contrast, the Leung \emph{et. al.} recovery is $F(\rho,\R\circ\E)\approx 1 -2.75 \gamma^2+\mathcal{O}(\gamma^3)$.

\subsection{Commentary on Examples}

It is not surprising that the optimum QER operation outperforms the 5 qubit code or the Leung code.  It is, however, noteworthy that the difference can be significant, for even relatively small values of $\gamma$.  We may conclude that channel specific recoveries can significantly improve the performance of error correcting procedures.  This conclusion was shared by Leung \emph{et. al.}, who lamented the lack of a general method to design such channel-specific schemes.  The SDP formalism outlined in this paper, provides such a general method.

Perhaps the most notable improvement in optimum QER can be noted in the `no recovery' comparisons in figs. \ref{fig:5qubit_comp} and \ref{fig:Leung_comp}.  These curves represent the entanglement fidelity of a single qubit transmitted through the noisy channel.  For values of $\gamma$ where the recovered entanglement fidelity lies below the `no recovery,' we see that the error mitigation procedure does more harm than good.  Performing optimum QER as opposed to QEC significantly extends the values of $\gamma$ for which error mitigation is valid.  This suggests optimum QER will be a particularly valuable technique for noisier systems.

Finally, it is worth noting the duality between optimum QER and optimum encoding.  We have derived the optimum operation $\R$ for a given encoding and noise process.  By the same process, one may derive the optimum encoding given a recovery operation and noise process.  This can most easily be seen by noting that the encoding operation is a valid quantum operation, and in fact, a spreading operation; it is thus subject to the same semidefinite cone constraints as the recovery operation.  In a manner similar to those suggested by \cite{Sho:03} and \cite{ReiWer:05}, one can conceivably obtain a channel-specific error recovery scheme by alternatively holding the recovery fixed and optimizing the encoding, and holding the encoding fixed and optimizing the recovery.  Full analysis of such a technique is deferred for future consideration.

\section{Conclusion}

The structure of quantum operations allow quantum error correction to be approached as an optimization problem.  Specifically, optimum recovery of an encoded quantum state from an error  process can be solved numerically using semidefinite programming when optimality is interpreted as a maximization of the entanglement fidelity.  This analysis suggests the ability to systematically search for recovery operations for complicated error schemes beyond those readily analyzed and corrected through more traditional QEC methodologies.  This problem is in general, the optimum combining operation following a quantum spreading channel, and thus a quantum parallel to the classical problem of diversity combining.

\begin{acknowledgements}
This work is sponsored by the Department of the Air Force
under AF Contract \#FA8721-05-C-0002. Opinions, interpretations, recommendations
and conclusions are those of the authors and are not necessarily
endorsed by the United States Government.
\end{acknowledgements}
\bibliography{quantum}

\end{document}